\documentclass[sigplan,10pt]{acmart}
\usepackage[utf8]{inputenc}
\usepackage{algorithm}
\usepackage{algorithmic}
\usepackage{fancyvrb}
\usepackage{amsmath}
\usepackage{pifont}
\VerbatimFootnotes
\usepackage{layouts}
\startPage{1}

\setcopyright{none}

\usepackage{booktabs}   
\usepackage{subcaption} 

\title{High Accuracy Low Precision QR Factorization and Least Square Solver on GPU with TensorCore}
\author{Shaoshuai Zhang, Panruo Wu}
\affiliation{Department of Computer Science\\University of Houston}
\email{{szhang36,pwu7}@uh.edu}
\date{July 2019}
\begin{document}
\begin{abstract}
Driven by the insatiable needs to process ever larger amount of data
with more complex models, modern computer processors and accelerators
are beginning to offer half precision floating point arithmetic
support, and extremely optimized special units such as NVIDIA TensorCore
on GPU and Google Tensor Processing Unit (TPU) that does half precision
matrix-matrix multiplication exceptionally efficiently.  In this paper
we present a large scale mixed precision linear least square solver that
achieves high accuracy using the low precision TensorCore GPU.  The mixed
precision system consists of both innovative algorithms and implementations,
and is shown to be up to 14$\times$ faster than single precision cuSOLVER at QR
matrix factorization at large scale with slightly lower accuracy, 
and up to 10$\times$ faster than double precision direct QR least square solver
with comparable accuracy. 

\end{abstract}
\maketitle

\section{Introduction}

Driven by the need to train large scale deep neural networks,
there's been a tidal wave of specialized low precision matrix
matrix multiplication units. Among them are TensorCore from
NVIDIA on its Volta and Turing architecture, Google's Tensor Processing
Unit (TPU)\footnote{\url{https://cloud.google.com/tpu/}},  and Intel's
upcoming Cooper Lake Xeon processors, as well as its Nervana
Neural Network Processor NNP-T 1000\footnote{\url{https://www.nextplatform.com/2019/07/15/intel-prepares-to-graft-googles-bfloat16-onto-processors/}}.  These specialized tensor
core units are usually characterized by the support of lower
precision arithmetic (such as 16 bit floating point FP16),
and extremely efficient matrix-matrix multiplication. For example,
NVIDIA V100 boasts 112Tera (112 trillion)
``deep learning'' FLOPS (floating point operation per second) \cite{Nvidia2017},
which is roughly half precision matrix multiplication accumulated
in single precision. Google's TPU v3 claims 420 TeraFLOPS,
also in doing half precision matrix-matrix multiplication.  In contrast,
V100 single precision peak performance is 14 TeraFLOPS, and double
precision is 7TeraFLOPS. Having these special units greatly speedups
the application that primarily spends time in low precision matrix-matrix
multiplication, and also results in much higher energy efficiency. 

However outside the neural network training and inference, effective use
of such tensor core units is much less well developed. There are two challenges: 
the application
must undertake primarily matrix-matrix multiplication, and it must have
stabilization procedures as half precision arithmetic is very limited
in accuracy and range. In this paper we present effective use of
NVIDIA TensorCore units to QR factorize matrix and solve linear least square
problem (LLS). QR factorization is a popular direct solver for
linear least square problem, and also a method for orthogonalization
of a set of vectors. 
Least square problem and its many variants are prevalent
in science, engineering, and statistical machine learning;
for instance non-linear least square problems 
are probably the largest source of all non-linear optimization problems. 
To give a specific example (gradiometry), consider the large scale least
square problems solved today concerning the determination of the Earth's gravity field from
highly accurate satellite measurements; see \cite{duff_parallel_2006}. Another example
is the least square problems arising
from many fields (data fitting, statistical machine learning, geodesy, computer
vision, robotics (bundle adjustment), etc). Non-linear least square problems
can often be solved as a series of linear least square problems. As such,
linear least square problem solvers form a core component of any linear algebra
packages such as LAPACK~\cite{Anderson1999} which have been downloaded millions
of times, and all major processor vendors provide architecture-optimized
reimplementations such as MKL from Intel, ACML from AMD, ESSL from IBM, 
cuBLAS/cuSOLVER from NVIDIA. 

Specifically, we develop a novel QR factorization that is able to exploit
TensorCore effectively to be 3x-14x faster on large scale than the NVIDIA
optimized cuSOLVER single precision QR subroutine at slightly lower accuracy.
To compensate the loss of accuracy, we combine the fast TensorCore QR
with a Krylov subspace iterative LLS solver to achieve high accuracy in a few iterations.
Here are the contributions of this paper. 
\begin{enumerate}
\item We propose a novel QR factorization algorithm that's designed to
exploit the emerging TensorCore technologies for speedup of up 2.9x-14.7x
on large scale matrices with a variety of shapes, 
compared to state-of-the-art cuSOLVER dense solver on NVIDIA GPU. 
\item We propose and demonstrate a novel combination of Krylov
Linear Least Square solver with low-precision QR factorization
to achieve single or double precision accuracy within a few iterations. Compared with \textbf{double precision} cuSOLVER, our
solution is usually more than 3x and up to 10x faster with comparable accuracy. 
\item We conduct comprehensive empirical study of the accuracy and performance
of QR factorization
and LLS hybrid solver for a variety of matrices, with different sizes,
aspect ratio, and spectrum distribution. 
\end{enumerate}
The paper is organized as follows. Section 2 introduces numerical, algorithmic,
and architectural backgrounds to understand this paper. Section 3 introduces
the main methods, analysis, and rationale behind our algorithmic design and
implementation. Section 4 is a comprehensive empirical study on the accuracy
and performance of the proposed methods. Section 5 discusses related work and
the context around this paper, and section 6 wraps up it with conclusion, limitations,
and future directions. 
\section{Backgrounds }
In this section we review some backgrounds that are most relevant to understand
this paper.  This is standard material; for readers already familiar with these topics
they are encouraged to quickly scan it. 
\subsection{Half Precision Arithmetic and TensorCore GPU}
NVIDIA introduced a specialized unit called TensorCore from their Volta architecture,
which boasts up to 112~TFLOPS ($112\times 10^{12}$ floating point operations per second)
for half precision (FP16) matrix-matrix multiplication.  Compared to single precision SGEMM
(Single precision GEneral Matrix-Matrix multiplication)
and double precision DGEMM, TensorCore is 7x and 14x faster respectively, which is
a considerable upgrade in the performance at the cost of significantly lower precision
and consequent loss of accuracy and numerical stability. 

TensorCore only supports matrix-matrix multiplication (GEMM\footnote{LAPACK subroutine naming convention: \verb|SGEMM| means single precision general matrix
multiplication, and \verb|DGEMM| means double precision one}).  The easiest to use API
is from cuBLAS, and it has many variations. A more flexible and also highly efficient
way to program TensorCore is through the CUTLASS template library\footnote{\url{https://github.com/NVIDIA/cutlass}} from NVIDIA, or directly
call the WMMA intrinsic. For this paper we use TensorCore through cuBLAS library. 


The Google Tensor Processing Unit (TPU) also depends extensively on 16 bits floating point
matrix-matrix multiplication to achieve its claimed 420 TFLOPS in its latest TPU v3 offering. 
However the 16 bits floating point format TPU uses is slightly different from the NVIDIA
TensorCore; TPU uses the bfloat16 format, which has 3 less bits for mantissa and use
3 more bits for exponents so it can represent a wider range of numbers at lower resolution. 
Intel also planned to introduce bfloat16 processing (together with FP32 accumulation)
in their future processors (Cooper Lake Xeon) so we will see more variety of half precision support in mainstream
processors, which makes it even more useful to extend the use pattern of low precision computing
beyond deep neural networks.

Let us take a look at the different floating point format and see what gives and what takes
in terms of accuracy (resolution in representing real numbers), and range (smallest and largest
representable real number):

\includegraphics[width=\columnwidth]{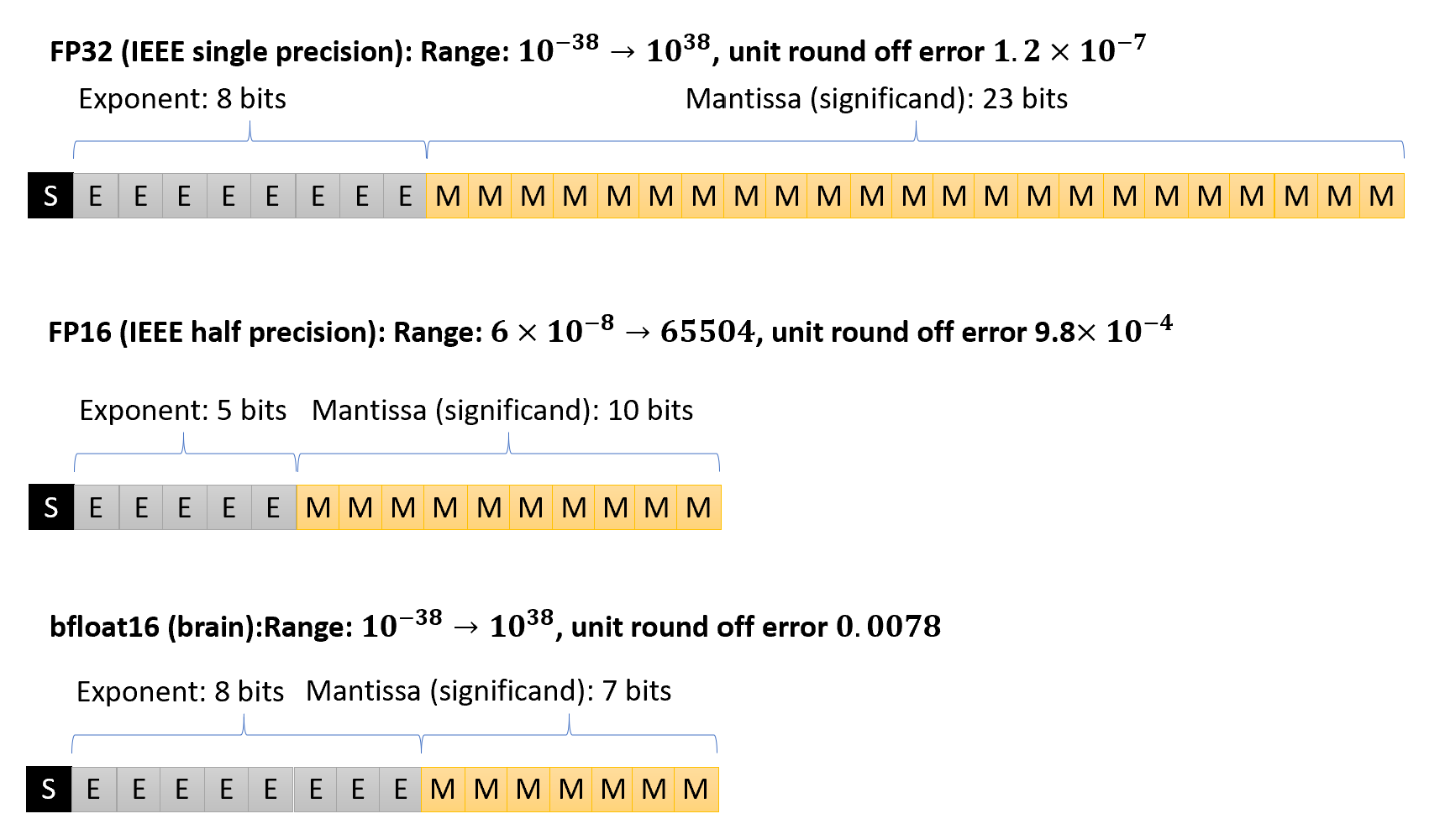} 
The IEEE single precision floating point format is accurate and widely ranged,
for it has 32 bits to spare.  There are currently two widely implemented 16 bits
floating point formats. 
Among them, IEEE FP16 has a significantly constrained range, but its resolution (the
unit roundoff error---the distance to next representable number from 1) is about 10 times
better than bfloat16. Bfloat16 on the other hand has the same range as IEEE FP32, but
its resolution is pitiful (there is no bfloat16 number between 1 and 1.0078). 
Thus bfloat16 is more robust (less prone to overflow and underflow) but less
stable/precise (large roundoff error).  In this paper we use FP16 format supported by
NVIDIA TensorCore. 


Error analysis of such low precision arithmetic is only emerging. In ~\cite{higham_simulating_2019} has error analysis that shows accumulating in higher precision
helps greatly in preserving accuracy in matrix/vector accumulation.

\subsection{Linear Least Square (LLS) Problems and Direct Solvers}
\label{subsec:lls_qr}
The (over-determined) linear least square problem is stated as an minimization problem:
\begin{equation}
\min_x ||Ax-b||^2\label{opt:lls}
\end{equation}
where $A\in \mathbb{R}^{m \times n} (m\ >n)$ has full column rank, and $b\in \mathbb{R}^n$.  Geometrically,
this minimization is to find the "projection" of point $b$ onto the range (column space) of
matrix $A$.  Analytically the LLS problem has closed form solution:
\begin{equation}
x^* = (A^TA)^{-1} A^Tb
\end{equation}
Computationally, the analytical solution can be obtained by solving the square linear equation (called the normal
equation): $ A^TAx = A^Tb$.  Typically a Cholesky factorization of $A^TA=LL^T$ can lead to a solution, via
backward and forward substitution. However directly forming $A^TA$ is unstable for all but the most well-conditioned systems; in practice we would avoid forming $A^TA$ directly.  Anyway this is our first
method: direct normal equation (NE) method:
\begin{equation}
A^TAx = A^Tb\label{eq:ne}
\end{equation}

The second direct method which can handle more ill-conditioned matrix is based on QR
factorization. For a tall and skinny matrix $A$ it takes roughly twice flops than the
NE method, but it handles a much wider range of matrix (if NE can handle up to 
condition number $\kappa$, then QR can handle condition number $\kappa^2$).
The basic idea is as follows. First we factorize the rectangular matrix 
$A\in \mathbb{R}^{m\times n}, m>n$ into
the product of an orthogonal matrix $Q\in \mathbb{R}^{m\times n}$, and
a square upper triangular matrix $R\in \mathbb{R}^{n\times n}$: $A=QR$. 
Then the solution to \eqref{opt:lls} is given by the following elementary matrix-vector
operation:
\begin{equation} 
x^* = R^{-1}(Q^Tb) \label{eq:qr_lls}
\end{equation}
which can be implemented as Algorithm~\ref{alg:lls_qr}.
\begin{algorithm}
\caption{Direct Linear Least Square Problem Solver with QR Factorization. The algorithm is in Matlab-like syntax.
The corresponding LAPACK subroutines are in the comments.}
\label{alg:lls_qr}
\begin{Verbatim}[numbers=left,xleftmargin=5mm,commandchars=\\\{\}]
[x] = function LLS_QR(A, b)
  [Q, R] = qr(A); % xGEQRF() 
  b = Q' * b;     % xORMQR()
  x = inv(R) * b; % xTRSM()
end
\end{Verbatim}
\end{algorithm}
For even more ill-conditioned system, or rank-deficient system (the columns of
$A$ are linearly dependent), we need more stable  and expensive algorithms such as rank-revealing
QR (e.g. QR with column pivoting), or Singular Value Decomposition (SVD). We do
not cover these types of methods, and confine ourselves to using the QR factorization
to solve modestly ill-conditioned LLS problem. 

\subsection{Iterative Solvers for LLS, and preconditioning}
As discussed in the previous subsection, direct solvers are robust but could be slow for
large scale problems.  Iterative methods are more attractive for large scale and especially 
sparse problems, where the only operation involving matrix $A$ is the matrix-vector
multiplication $Av$ and $A^Tv$.  However for iterative methods to be competitive
a good preconditioner is essential, which is in general a very difficult problem. 
A basic algorithm for solving the LLS problem without explicitly forming $A^TA$ is
called CGLS\footnote{this algorithm has been given various names, such as
CGNR, CGNE, and GCG-LS}.  Basically CGLS amounts to applying the famous
conjugate gradient (CG) method on the normal equation, without explicitly forming
$A^TA$, thus avoiding squaring the condition number. 

In this paper, we are going to combine the direct solver based on QR factorization,
with an iterative as safeguards to refine accuracy (this idea may be broadly called iterative
refinement).  The hope is that we can get the best of both worlds---the opportunity to
use TensorCore and predictability/stability of direct methods, and flexibility to take
an inaccurate solution/factorization and turn it into increasingly accurate solution
through iteration.

\section{Methods }
In this section we describe our TensorCore accelerated QR factorization first,
and the use of iterative refinement to refine the accuracy of LLS solutions
based on the TensorCore QR.

\subsection{The TensorCore Accelerated QR Factorization }
As briefly introduced in section \ref{subsec:lls_qr}, QR factorization is
one of the most fundamental matrix factorization in numerical linear algebra.
It seeks to factorize a general matrix $A$ into product of an orthogonal matrix
$Q$ and an upper triangular matrix $R$. The use of QR factorization includes
solving linear least square problem, and orthogonalization of columns of $A$
(columns of $Q$ are a orthonormal basis for the column space of $A$, or the range
of $A$), and in singular value computation. 
As such, QR is almost always an important building block of any numerical
linear algebra packages such as LAPACK~\cite{Anderson1999}, ScaLAPACK\cite{Blackford1997}.
On GPU, NVIDIA provides well optimized cuBLAS for basic matrix operations such as
multiplication, and cuSOLVER for high level matrix factorizations, such as LU/QR and
eigendecompositions. A more comprehensive package is the MAGMA~\cite{Dongarra2014a},
which uses a hybrid CPU/GPU architecture. 

In the following subsections we will describe our three attempts to speedup QR
factorization on GPU with TensorCore, with the first obvious one but failed to produce
speedup, and a mildly successful second one, and the third
reasonably good attempt. 

\subsubsection{First Attempt: Replacing GEMM with TensorCore GEMM}


Unlike matrix-matrix multiplication, matrix factorization typically 
exhibits more dependency and
less parallelism,
and more complicated memory access pattern.  Therefore matrix factorization cannot
achieve the speed of matrix-matrix computation, but with algorithmic innovations
called "blocking" or "tiling" could approach a significant fraction of it. Basically, the idea
of tiling is to aggregate matrix-vector operations into fewer but bigger matrix-matrix
multiplications, so as to increase arithmetic intensity (ratio between operations and
number of elements) therefore enabling better data reuse. This technique is essential
in bridging the gap between fast processor and slow memory, using the fast on-chip
memories (registers, caches) to service most of the memory access.  But because
of the complex dependency, some part of the factorization is still matrix-vector and
vector-vector based, which are much slower than matrix-matrix operations.  Modern
algorithms and implementations usually divides each iteration of the factorization
into two steps: \textbf{panel factorization} (slow, but small) and \textbf{trailing matrix update} (fast and big), where
most the floating point arithmetic are spent in the trailing matrix update. 

Based on this structure, the first attempt keeps the panel factorization intact,
while replacing the trailing matrix update with TensorCore GEMM. This strategy
is simple but turns out to be naive.  MAGMA~\cite{Dongarra2014a} QR uses hybrid
CPU/GPU architecture where panel factorization is on CPU, and trailing matrix
(big GEMM) is on GPU. Due to algorithmic pipeline, the GEMM execution is completely
overlapped by the panel, thus speeding up GEMM has no effect on the overall QR
speed. NVIDIA cuSOLVER is a pure GPU implementation, and we can use cuSOLVER
QR as the panel, and cuBLAS GEMM with TensorCore for trailing matrix.
But unfortunately this
results in \textit{speed down}  than cuSOLVER QR, for reasons unknown
to us (cuSOLVER is not open source).

To summarize, our first obvious attempt that tries to flip a switch to replace every occurrence of matrix-matrix multiplication with TensorCore accelerated version
does not result in speedup, for both CPU/GPU hybrid QR and GPU native QR. 
\subsubsection{Second Attempt: Recursive Gram-Schmidt QR Factorization}
There is another variant of QR algorithm that can also turn most of its operations
into matrix-matrix multiplication---recursive QR. The idea of recursive QR has been
explored by \cite{elmroth_applying_2000} to replace the \textbf{panel} factorization
in QR. It's only used in panel because it
increases the number of operations needed to 2x that of Householder QR. The
big increase in operation counts is probably the reason that recursive QR is
not used often in practice. On the other hand, Recursive QR has the advantage of increased data locality,
thus the limited use of QR in panel factorization is able to balance out its
limited increased operation count, and get modest overall speedup. 

Our second attempt is going to take the recursive QR as
the overall QR algorithm, and use the cuSOLVER QR when the recursion becomes
sufficiently small (panel).  We mitigate the increase of operations, by resorting to a different
basic QR algorithm---(modified) Gram-Schmidt (MGS)---rather than conventional Householder
QR.  It turns out that with MGS Recursive QR, the operation counts only increases
moderately compared to Householder QR ($2mn^2$ vs $2mn^2-\frac23 n^3$), 
instead of two times increase. But because we can dramatically acclerate the
matrix-matrix multiplication, it has the potential to result in faster overall execution time. 

The basic idea of recursive QR is a quite simple one.  Given a matrix $A$, we divide evenly its columns
into two halves, denoted by $A=[A_1 |  A_2]$. We first QR factorize the first half $A_1=Q_1R_{11}$,
and then compute north-east quarter of $R_{12}=Q_1^T A_2$. Next we update the
second half $A_2 = A_2 - Q_1 R_{12}$. Finally QR factorize the updated second half
$A_2 = Q_2 R_{22}$. Note that the QR of the two halves can be recursed using this
algorithm itself.  The result of the original QR factors can be assembled like this:
\begin{equation}
[A_1 | A_2 ] = [Q_1 | Q_2 ] \left[ \begin{array}{c|c}
R_{11} & R_{12} \\ \hline
 & R_{22}
\end{array}   \right]\label{eq:recqr}
\end{equation}
\begin{algorithm}
\caption{Recursive Modified Gram-Schmidt QR Factorization,
with recursion cutoff size 128}\label{alg:mgs_qr}
\begin{Verbatim}[numbers=left,xleftmargin=5mm,commandchars=\\\{\}]
function [Q,R] = \textcolor{blue}{RMGSQR}(A)
  [m,n] = size(A);
  if n==128
    [Q,R] = panelQR(A);
    return
  end
  [Q1,R11] = \textcolor{blue}{RMGSQR}(A(:,1:n/2);
  R12 = Q1' * A(:,n/2+1:n);
  [Q2,R22] = \textcolor{blue}{RMGSQR}(A(:,n/2+1:n) - Q1 * R12);
  Q = [Q1 Q2];
  R = [R11 R12; zeros(n/2) R22];
end
\end{Verbatim}
\end{algorithm}

\begin{algorithm}
\caption{Recursive tiled HOUseholder QR Factorization,
with recursion cutoff size 128}\label{alg:house_qr}
\begin{Verbatim}[numbers=left,xleftmargin=5mm,commandchars=\\\{\}]
function [Y,T,R] = \textcolor{blue}{RHOUQR}(A)
  [m,n] = size(A);
  if n==128
    [Y,T,R] = panelQR(A);
    return
  end
  [Y1,T1,R1] = \textcolor{blue}{RHOUQR}(A(:,1:n/2);
  B=A(:,n/2+1:n)-(Y1*T1')*(Y1'*A(1:m,n/2+1:n));
  [Y2,T2,R2] = \textcolor{blue}{RHOUQR}(B(n/2+1:m,:));
  R = [R1 B(1:n/2,:); zeros(n/2) R2];
  Y = [Y1,  [zeros(n/2); Y2] ] ;
  T = [T1, -T1*(Y1'*Y2)*T2; zeros(n/2), T2];
end
\end{Verbatim}
\end{algorithm}

Here is the contrast between Recursive Householder QR (Algorithm~\ref{alg:house_qr}) and Recursive MGS QR (Algorithm~\ref{alg:mgs_qr})
in matlab-like syntax\footnote{To read the algorithms: \verb|A(i:j,l:k)| denotes the
submatrix of \verb|A| with the i to j-th row and l to k-th columns; \verb|[A B]| or \verb|[A,B]|
returns the horizontal concatenation of matrix \verb|A,B| with the same number of rows;
\verb|[A; B]| is the vertical concatenation; \verb|A'| is the transpose. }
Note that Algorithm~\ref{alg:mgs_qr} follows more closely the recursion~\eqref{eq:recqr}
while Algorithm~\ref{alg:house_qr} deviates slightly. In contrast, Algorithm~\ref{alg:house_qr}
looks more complicated, and it does \textit{more} operations, primarily due to the
need to reconstruct the blocks \verb|T,Y| in line 11 and 12. This is due to the implicit
representation of the orthogonal factor $Q$ as Householder reflectors in Householder QR
algorithm; see~\cite{schreiber_storage-efficient_1989}. 

Now we can complete our second attempt.  The basic structure is the Algorithm~\ref{alg:mgs_qr},
and the implementation uses cuSOLVER \verb|SGEQRF()|  as the \verb|panelQR| (line 4) when
the input matrix $A$ becomes small ($n=128$). For matrix size $m \times n$, this algorithm
roughly takes $2mn^2$ flops.  In each function call \verb|RMGSQR()|, roughly half
of the flops is in matrix-matrix multiplication as shown in line 8 and 9 (in parenthesis),
and the other half of the flops spent in the two recursion function calls. We use
TensorCore to accelerate these matrix-matrix multiplications. The resulting implementation
is up to 1.4x faster than the NVIDIA cuSOLVER \verb|SGEQRF()| subroutine
for matrix size $32768\times 16384$. This is a step forward from the first attempt;
we keep using the cuSOLVER QR as our panel, and devised a different
QR algorithm based on recursive Gram-Schmidt instead of tiled Householder algorithm.
These changes enable TensorCore to accelerate the overall performance of
QR factorization. In the next subsection, we are going to replace the cuSOLVER QR
panel with a faster one, such that the potential of TensorCore is further revealed.
\subsubsection{Further Optimization: Communication Avoiding Panel}
The second attempt is encouraging, but profiling shows that most of 
the time (>\%80) is spent in the \verb|panelQR|, even though the panel
only constitutes a small fraction of operations. The matrix-matrix multiplication
is simply too fast, which just exposes the panel as dominating bottleneck.
\textbf{Thus to really unlock the speed of TensorCore, we
need a much faster} \verb|panelQR|;  the cuSOLVER \verb|SGEQRF| is
taking so much time that that accelerating the other matrix-matrix multiplication
reduces execution time only marginally. 

The challenge in fast \verb|panelQR| is that of data locality and 
parallelism.  The conventional Householder panel has sequentially
dependent iterations, and the working-set is the whole panel
which cannot fit in fast memory on GPU (register files+ shared memory). 
Fortunately for QR, there's a communication avoiding QR (CAQR)~\cite{Anderson1999} 
variant that simultaneously improve parallelism and data locality. 
Our \verb|panelQR| is based on CAQR, with the Modified Gram Schmidt
QR replacing Householder QR used in~\cite{Anderson1999}. 
The idea of CAQR can be illustrated in the following equation:

In \eqref{eq:caqr}, there are 5 steps indicated by the number over the equality sign. In the \ding{172}
step, we divide a tall matrix $A$ evenly into 4 smaller matrix (still tall, more rows than columns),
and QR factorize them independently.  In step \ding{173} we stack the R factors vertically.  Note that
the number of rows of the R factors are less than the number of rows of original $A$. In step
\ding{174}, we factorize the vertically stacked $R$s (potentially carry this process recursively). 
In \ding{175}, we do 4 matrix-matrix multiplications for the 4 corresponding $Q$ factors. 
In \ding{176} we reinterpret the result as the QR factors of original $A$. The reason $Q$
is orthogonal, is that in step \ding{175} the 4 matrix-matrix multiplication is equivalent
to the product of two orthogonal matrices (second line), and therefore is orthogonal. 
\begin{equation}
\begin{aligned}
\left[\begin{array}{c}
A_1 \\ 
A_2 \\ 
A_3 \\ 
A_4
\end{array}\right] &\overset{\text{\ding{172}}}{=} \left[\begin{array}{c}
Q_{11}R_1 \\ 
Q_{12}R_2 \\ 
Q_{13}R_3 \\ 
Q_{14}R_4
\end{array} \right] \overset{\text{\ding{173}}}{=} \left[\begin{array}{cccc}
Q_{11} &  &  &  \\ 
 & Q_{12} &  &  \\ 
 &  & Q_{13} &  \\ 
 &  &  & Q_{14}
\end{array} \right] \left[ \begin{array}{c} R_1 \\R_2 \\ R_3\\R_4\end{array}\right]\\
&\overset{\text{\ding{174}}}{=}\left[\begin{array}{cccc}
Q_{11} &  &  &  \\ 
 & Q_{12} &  &  \\ 
 &  & Q_{13} &  \\ 
 &  &  & Q_{14}
\end{array} \right] \left[ \begin{array}{c} Q_{21} \\Q_{22} \\ Q_{23}\\Q_{24}\end{array}\right] R\\
&\overset{\text{\ding{175}}}{=}  \left[ \begin{array}{c} Q_{11}Q_{21} \\Q_{12}Q_{22} \\ Q_{13}Q_{23}\\Q_{14}Q_{24}\end{array}\right]R \overset{\text{\ding{176}}}{=} QR
\end{aligned}\label{eq:caqr}
\end{equation}
Practically, we fix our panel to be of 32 columns with $m$ rows, and decompose
the matrix $A$ into 256x32 submatrices (step \ding{172}). On V100 GPU, the
256x32 submatrix can fit into shared memory so that we only need to read
and write global memory once. These 256x32 blocks are independently
factorized using the modified Gram-Schmidt algorithm into QR factors; see
algorithm~\ref{alg:getrf32}. To map this algorithm to GPU, we let each threadblock
QR factorize one 256x32 block. The implementation of Algorithm~\ref{alg:getrf32}
within a threadblock is straightforward. We launch 256 threads, with each threads
reading,processing, and writing a single row of the 256x32 block. The most time consuming
part is line 7 where reductions are needed (vector inner products across threads). 
We use CUB template library\footnote{\url{https://nvlabs.github.io/cub/}} from NVIDIA Research to have a threadblock level fast reduction.
We manually unroll the loop 4 ways to expose more instruction level parallelism,
and to reduce the number of reductions by a factor of 4. In step \ding{175} we use
cuBLAS batched \verb|SGEMM()| subroutine to do the matrix multiplications in
parallel. We recurse in step \ding{174}, until the number of rows is below 256 so that
a single threadblock will suffice.  In summary, our CAQR implementation has two salient
features: 1) the Gram-Schmidt process is run completely within shared memory; 
2) all the inter-threadblock communication/synchronization happens in the
batched SGEMM() which is extremely fast. Hence our CAQR panel reads global
memory minimally ($\log_8 (m/256) $ passes to the panel) , 
and have minimal cross threadblock synchronization and
communication.

\begin{algorithm}
\caption{256x32 Modified Gram-Schmidt QR}\label{alg:getrf32}
\begin{Verbatim}[numbers=left,xleftmargin=5mm,commandchars=\\\{\}]
function [Q,R] = mgs(A)
  [m,n] = size(A);
  Q = A; R = zeros(n);
  for k=1:n
    R(k,k) = norm(Q(:,k));
    Q(:,k) = Q(:,k)/R(k,k);
    R(k,k+1:n) = Q(:,k)' * Q(:,k+1:n);
    Q(:,k+1:n) = Q(:,k+1:n) - Q(:,k) * R(k,k+1:n)
  end
end
\end{Verbatim}
\end{algorithm}

The effect of our hand coded CAQR panel replacing the
\verb|panelQR| in Algorithm~\ref{alg:mgs_qr}
results in substantial speedup
over cuSOLVER \verb|SGEQRF()|. 
See figure~\ref{fig:qr_perf} for more details. 

%
%
%
%

\subsection{Linear Least Square Problem With QR Factorization}
One important use of QR factorization is to solve linear least square
problems.

\subsubsection{Numerical Issues}

A natural concern for using the half precision TensorCore matrix-matrix multiplication
is the potential loss of accuracy and stability.  In the case of QR, two kinds of accuracy
are of importance: the backward error and the orthogonality of the Q factor.  The backward
error is 
$$\frac{||A-\hat{Q}\hat{R}||_2}{||A||_2}$$
and the orthogonality of $\hat{Q}$ is
$$ ||I-\hat{Q}^T \hat{Q}||_2$$

The Recursive MGS QR has the property that the backward error is always quite small
(up to the working accuracy)
regardless of the conditioning of the matrix $A$, but the orthogonality loss bound is proportional
to the condition number of $A$; see \cite{bjorck_solving_1967}. This may limit on the
range of matrix $A$ that can be usefully factorized by our TensorCore Recursive MGS QR.
Specifically, when the matrix $A$ is too ill-conditioned, the Recursive MGS QR may lead
to unorthogonal $Q$.  We will
revisit this issue empirically in the experiment section later.

\subsubsection{Direct Solve with QR}
The accuracy of direct solution of LLS problem using QR factorization using \eqref{eq:qr_lls}
depend on the accuracy of the $QR$ factorization. To measure the accuracy of a solution to
the linear least square problem $\min_x ||Ax-b||$, we use the following accuracy metric:
$$ A^T (A\hat{x}-b) $$ 
for a computed solution $\hat{x}$.  Ideally this metric should be 0,
but will not be exactly zero due to roundoff errors in the QR factorization.  
Therefore smaller is better for this accuracy test for LLS.

\subsubsection{Iterative Refinement}
It can be seen that directly solve the LLS problem with our low precision QR factorization
may not lead to sufficient accuracy.  To achieve higher accuracy
we can refine the solution to get higher
accuracy.  There are two approaches for this task. One is actually called iterative refinement
in the literature~\cite{bjorck_iterative_1967,bjorck_iterative_1968,Higham2002,Demmel2007}.  Another one,
which appears to be new for this purpose is what we are going to introduce.  It's a Krylov
subspace iterative solver for LLS, coupled with our low-precision QR factorization as \textbf{preconditioner}
to achieve high accuracy and fast convergence. This idea blurs the distinction between direct
solver and iterative solver; it inherits the stability and robustness of direct solver, while retains
the flexibility and the iterative nature of Krylov iterative solver.  We use the CGLS iterative solver,
which is mathematically equivalent to Conjugate Gradient on the normal equation, but numerically
more stable. We list the algorithm with the QR factorization in Algorithm~\ref{alg:cgls_qr}.

\begin{algorithm}
\caption{LLS High Accuracy Solver: CGLS with RMGSQR as Preconditioner\textsuperscript{a}}\label{alg:cgls_qr}
\begin{Verbatim}[numbers=left,xleftmargin=5mm,commandchars=\\\{\}]
function [x] = cgls_qr(A,b)
  [Q,R] = RMGSQR(A); % TensorCore 
                     % Accelerated QR
  [m,n] = size(A);
  x = zeros(n,1);
  r = b - A*x;
  s = A'*r;
  p = s;
  norms0 = norm(s);
  gamma  = norms0^2;
  for k=1,2,...
    q      = A*(inv(R)*p); % preconditioned
                           % by R
    delta  = norm(q)^2;
    alpha  = gamma/gamma1;
    x      = x + alpha*p;
    r      = r - alpha*q;
    s      = inv(R')*(A'*r); % preconditioned
                             % by R
    norms  = norm(s);
    gamma1 = gamma;
    gamma  = norms^2;
    beta   = gamma / gamma1;
    p      = s + beta*p;
  end
end
\end{Verbatim}
{\small\textsuperscript{a} The convergence test is omitted.  This presentation is adapted from Per Christian Hansen and Michael Saunders at \url{https://web.stanford.edu/group/SOL/software/cgls/matlab/cgls.m}}
\end{algorithm}

This algorithm first calls upon the fast Recursive MGS QR to do QR factorization,
and then runs CGLS algorithm, with the R factor as right preconditioner for $A$. 
For a sufficiently accurate QR factor $R$, $AR^{-1}$ should be fairly well-conditioned,
which means that $\kappa(AR^{-1})$ is small (close to 1, ideally).  
The convergence rate is linear; specifically the error is
reduced by at least a constant factor in every iteration:
$$ e_k = e_0\left(\frac{\kappa(AR^{-1})-1}{\kappa(AR^{-1})+1}\right)^k $$
With perfect QR factorization $\kappa(AR^{-1})=\kappa(Q)=1$, and CGLS converges in
1 iteration. With imperfect QR, we need slightly more iterations to converge; see
experiment section \ref{subsec:lls_perf} for some empirical examples. 

\section{Experiments}
In this section we conduct comprehensive empirical study on the
numerical behavior (accuracy), and performance behavior of our proposed
Recursive MGS QR factorization and Linear Least Square Solver.

For all the experiments we use a Redhat 7 Linux workstation
with NVIDIA V100 (PCIe version) GPU.  The CUDA version is 10.1,
which contains a C++ compiler and libraries cuBLAS and cuSOLVER.
For the Linear Least Square experiments we used random matrix
generation routine from MAGMA 2.5.1 to generate random matrix
with specific condition number and singular value distribution. 

\subsection{QR Factorization}
\subsubsection{Performance}
\begin{figure}
\includegraphics[width=1.0\columnwidth]{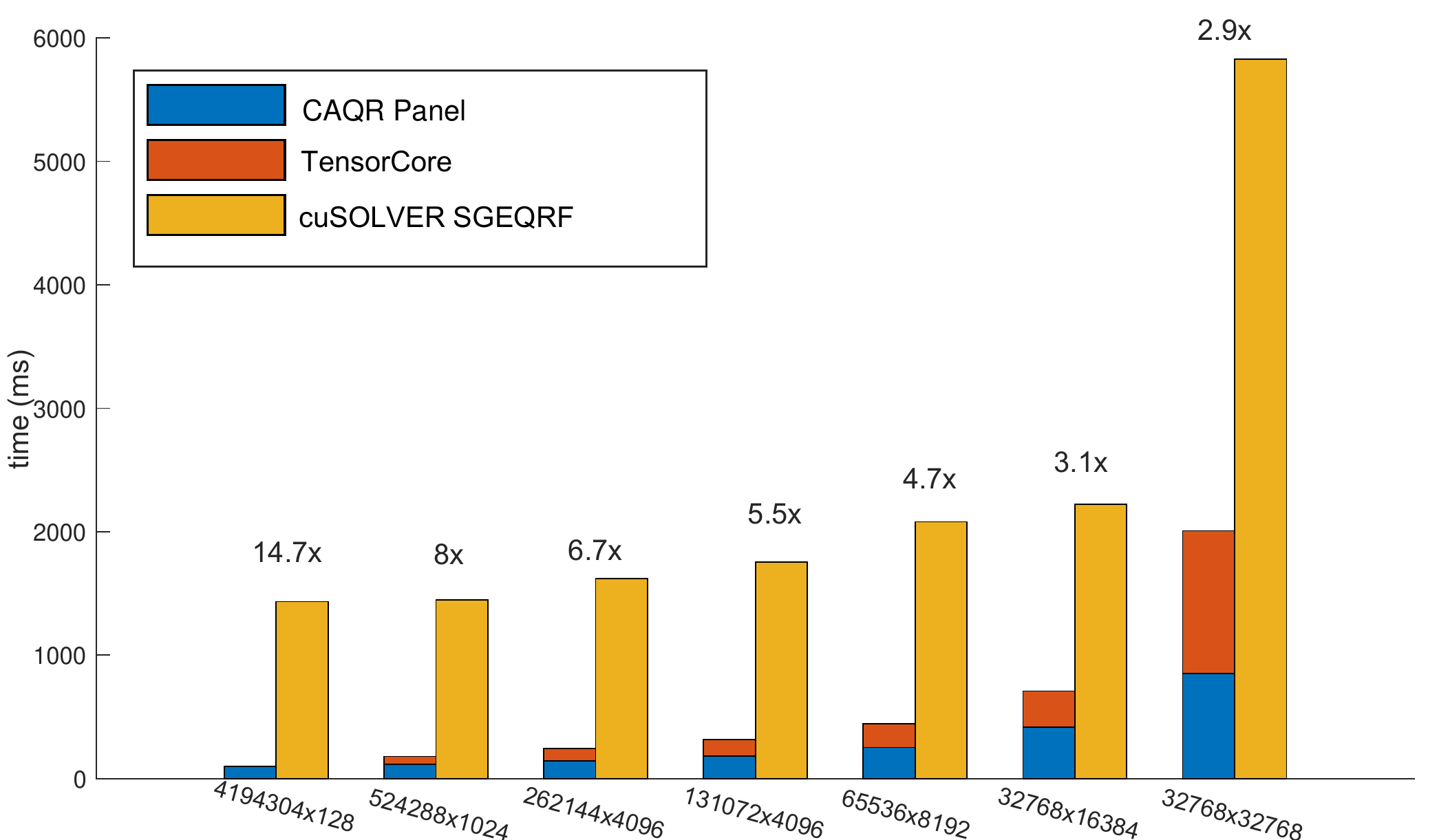} 
\caption{QR factorization performance: RMGSQR (left bar) vs. cuSOLVER SGEQRF (right bar)
for different matrix sizes. }\label{fig:qr_perf}
\end{figure}

Figure~\ref{fig:qr_perf} shows the performance of Recursive MGS QR in comparison
with the NVIDIA optimized cuSOLVER SGEQRF(). As we can see that for large scale
matrix, the speedup of TensorCore accelerated RMGSQR is between 2.9x to 14.7x,
depending on the shape of the matrix. Typically, the more tall and skinny the matrix
is, the higher speedup. For a square matrix the speedup is at its lowest 2.9x, and
for an extremely tall and skinny matrix (4194304$\times$128) the speedup is 14.7x.
Generally speaking the speedup of RMGSQR over SGEQRF is robust across board. 

\subsubsection{Accuracy}
\begin{figure}
    \centering
    \includegraphics[width=\columnwidth]{{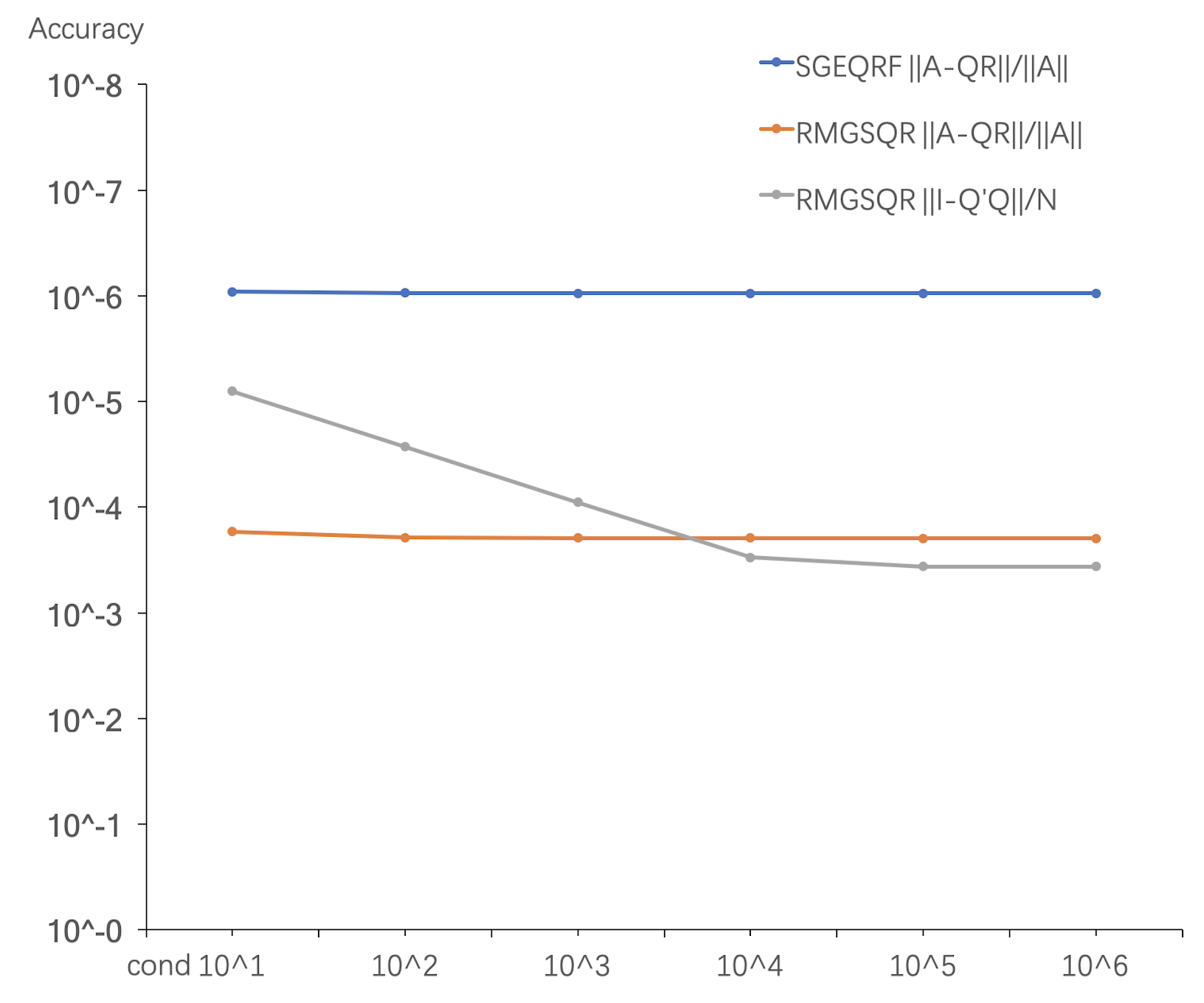}}
    \caption{QR factorization accuracy: RMGSQR $\frac{||A-QR||}{||A||}$ vs. SGEQRF$\frac{||A-QR||}{||A||}$ and RMGSQR $\frac{||I-Q^TQ||}{N}$, matrix size 8192*4096, SVD arithmetic distribution}
    \label{fig:qr_accu}
\end{figure}

As RMGSQR involves in half precision and because of rounding errors, we are not anticipating the same level accuracy when compared with cuSolver \verb SGEQRF and \verb DGEQRF. On the one hand, we can observe in Figure~\ref{fig:qr_accu} that the backward error  $\frac{||A-QR||}{||A||}$ of both RMGSQR and SGEQRF remains at a stable level, but the results of SGEQRF are more accurate than RMGSQR; On the other hand, normalized $||I-Q^TQ||/N$, which represents the orthogonality of the $Q$ factor, deteriorates as condition number increases,
but seems to stablize after cond $10^4$. 

This loss of orthogonality may be a problem or not, depending
on what QR is used for.  For solving LLS problem, direct solve
based on QR does not seem to suffer from the loss of 
orthogonality by much; see Figure~\ref{fig:lls_accu}
for example. For solving LLS problem iteratively as in 
Algorithm~\ref{alg:cgls_qr}, we are not using the Q factor,
and it's unclear whether the loss of Q orthogonality
is a problem or not; we seem to get pretty good results
in most cases (see the next subsection).
The difficulty depends more on
the distribution of singular values rather than the
condition number itself (thus the loss of orthogonality). 
For orthogonalization of a set of vectors using QR, the loss of orthogonalization
could be a problem for ill-conditioning. 
One immediate remedy is to \textit{re-orthogonalize},
namely taking a second QR factorization of the factor $Q$ itself.
This will remove the loss of orthogonality by large condition
number, at the cost of doubling the execution time.

{}

\subsection{Linear Least Square Problem}\label{subsec:lls_perf}
Unlike QR, whose accuracy only depends on condition number,
to refine LLS solution the CGLS iterative solver performance
depends on the singular value distribution of $A$.  To cover a comprehensive
variety of different singular value distribution and condition number,
we use the following randomly generated matrix. 1) each element is
i.i.d. from uniformly distributed random number within (0,1) and (-1,1);  2) each element is
i.i.d from normally distributed random number with mean 0 and standard
deviation 1; 3) random matrix with specified condition number and
geometric singular values ($\sigma_i$) distribution: 
$[\log\sigma_1,\ldots,\log\sigma_n]$ are evenly spaced; 4) random matrix with specified condition number and
arithmetic singular values ($\sigma_i$) distribution: 
$[\sigma_1,\ldots,\sigma_n]$ are evenly spaced; 5) random matrix
with clustered singular values: all but the smallest singular values
are 1. 

\subsubsection{Performance}
\begin{figure*}
    \centering
    \begin{subfigure}[b]{1.0\columnwidth}
        \includegraphics[width=\textwidth]{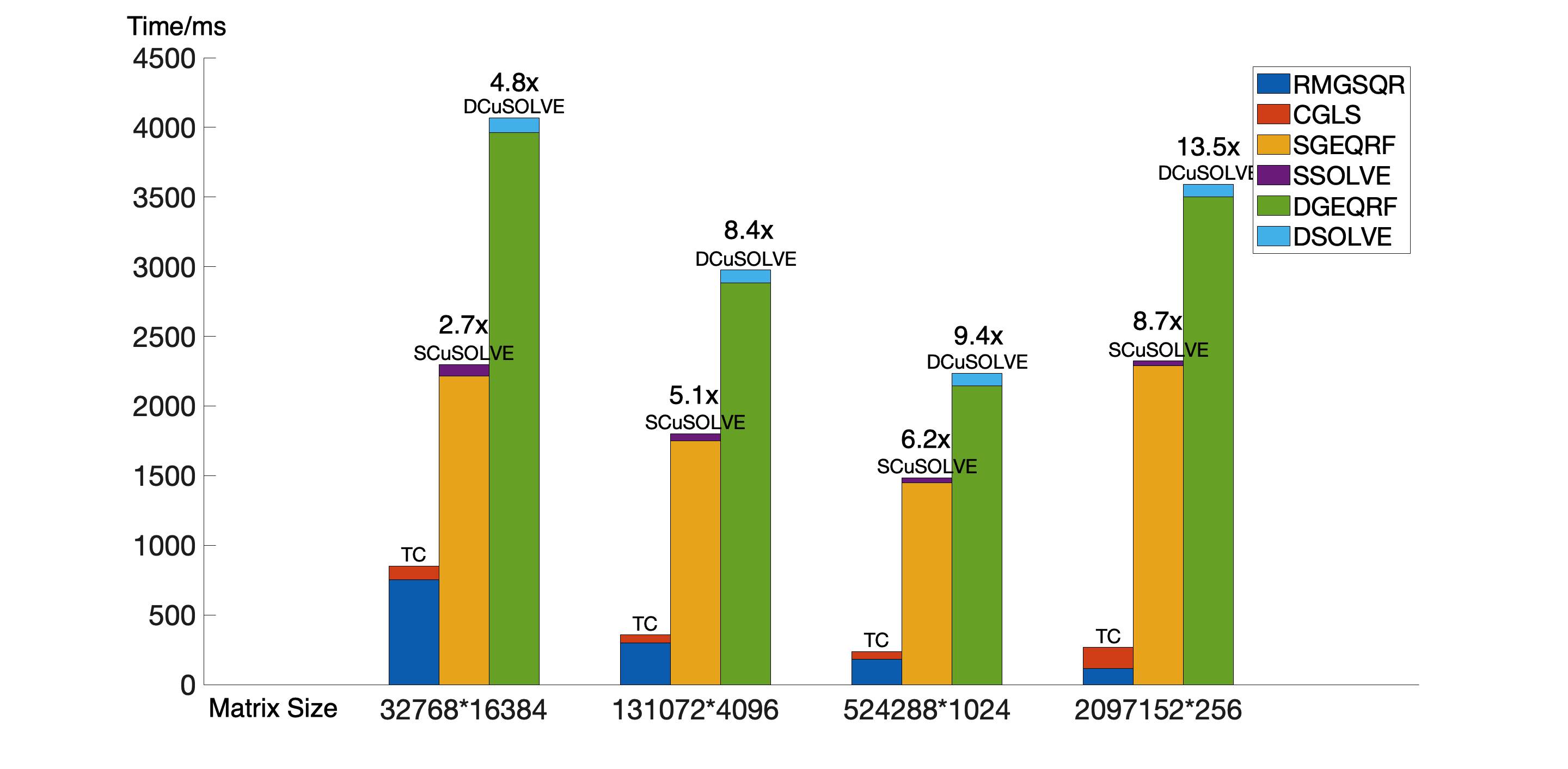}
        \caption{\normalsize Matrix of Type 1: Random Uniform on (0, 1)}
        \label{fig:3(a)}
    \end{subfigure}
    ~ 
    \begin{subfigure}[b]{1.0\columnwidth}
        \includegraphics[width=\textwidth]{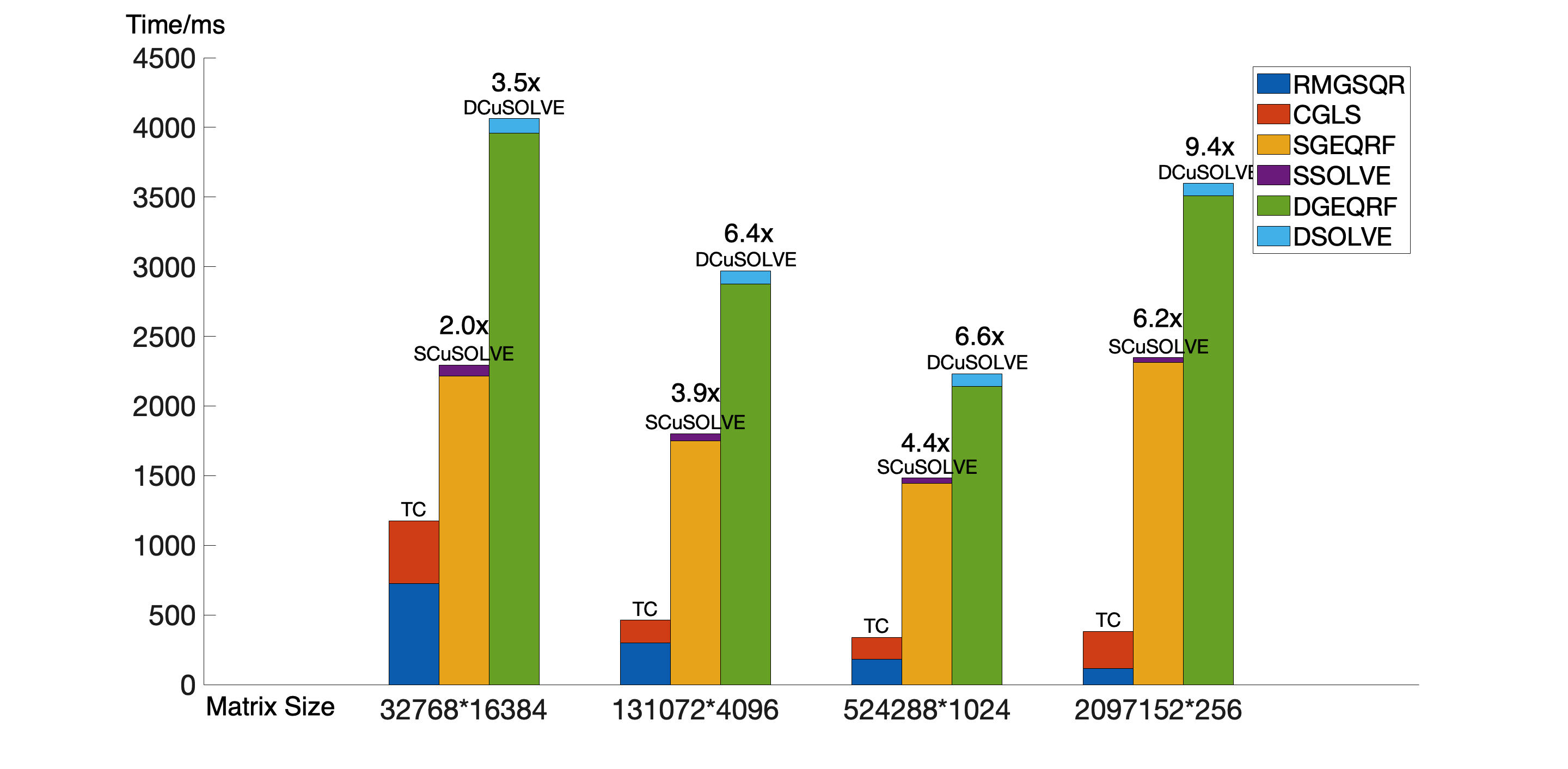}
        \caption{\normalsize Matrix of Type 2: random uniform on (-1, 1)}
        \label{fig:3(b)}
    \end{subfigure}
    
    \begin{subfigure}[b]{1.0\columnwidth}
        \includegraphics[width=\textwidth]{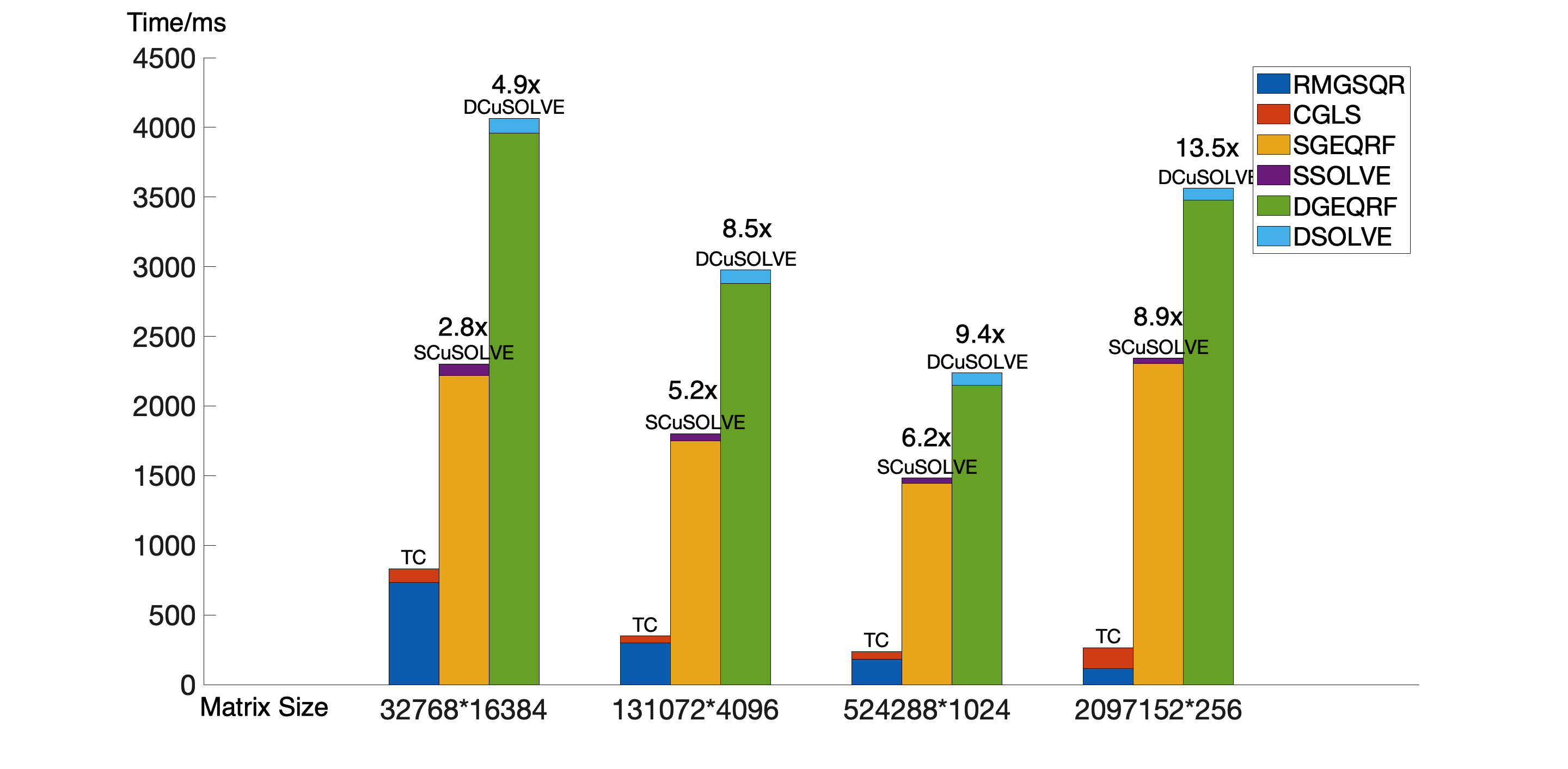}
        \caption{\normalsize Matrix of Type 3: random normal with mean=$0$,  
standard deviation=$1$}
        \label{fig:3(c)}
    \end{subfigure}
    \begin{subfigure}[b]{1.0\columnwidth}
        \includegraphics[width=\textwidth]{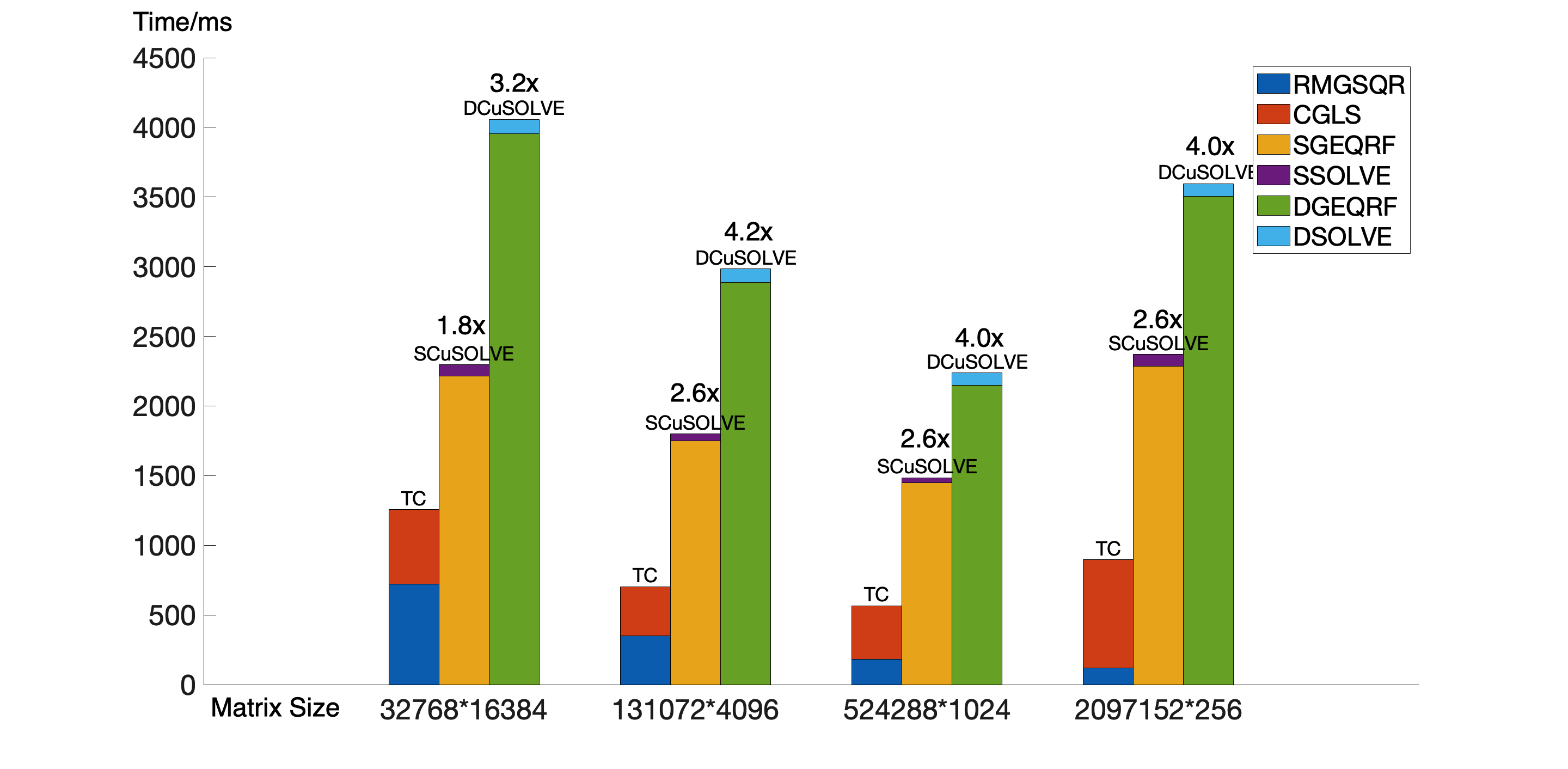}
        \caption{\normalsize Matrix of Type 4: SVD geometric distribution with $cond=10^3$}
        \label{fig:3(d)}
    \end{subfigure}
    
    \begin{subfigure}[b]{1.0\columnwidth}
        \includegraphics[width=\textwidth]{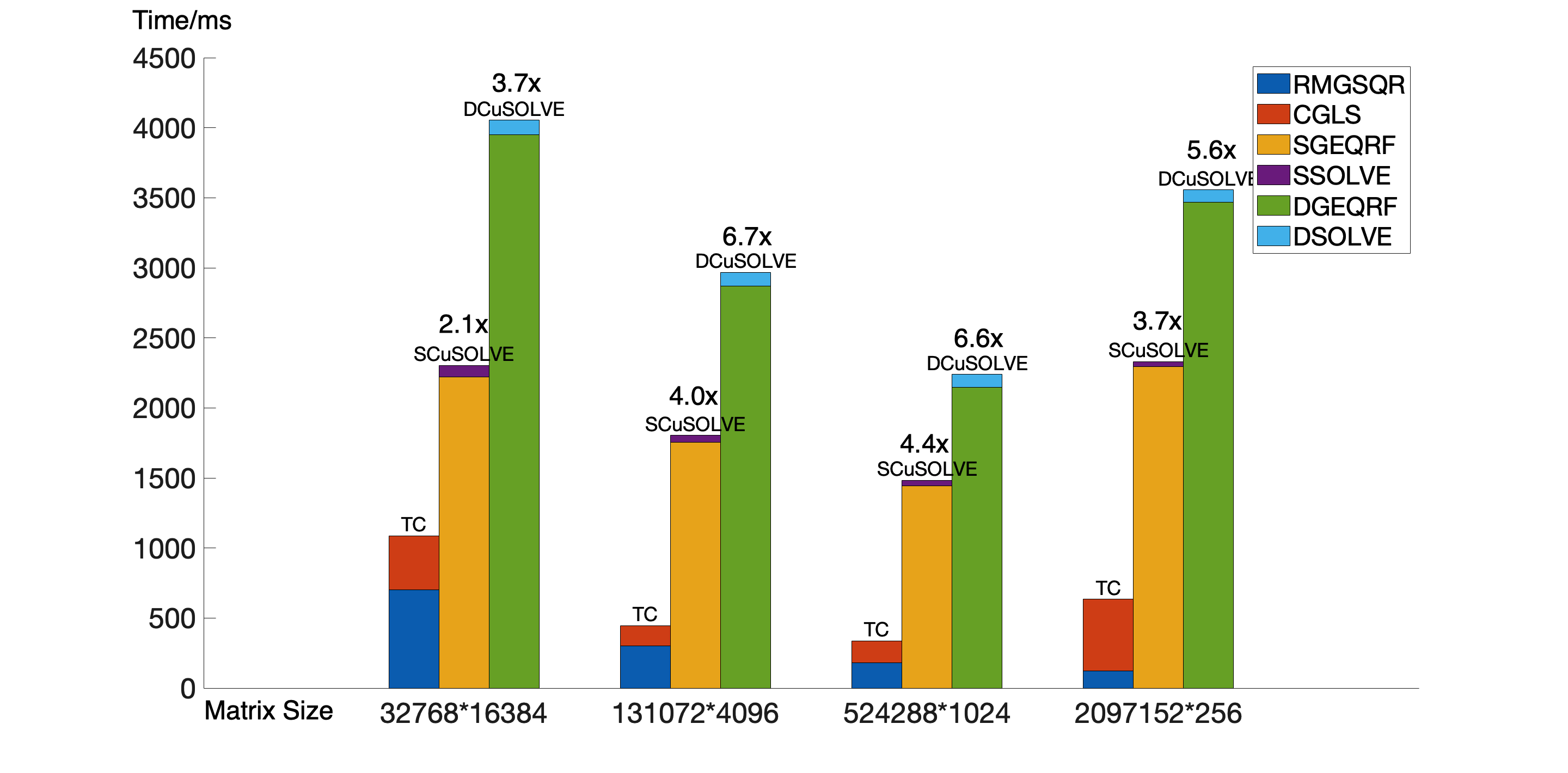}
        \caption{\normalsize Matrix of Type 5: SVD arithmetic distribution with $cond=10^5$}
        \label{fig:3(e)}
    \end{subfigure}
    \begin{subfigure}[b]{1.0\columnwidth}
        \includegraphics[width=\textwidth]{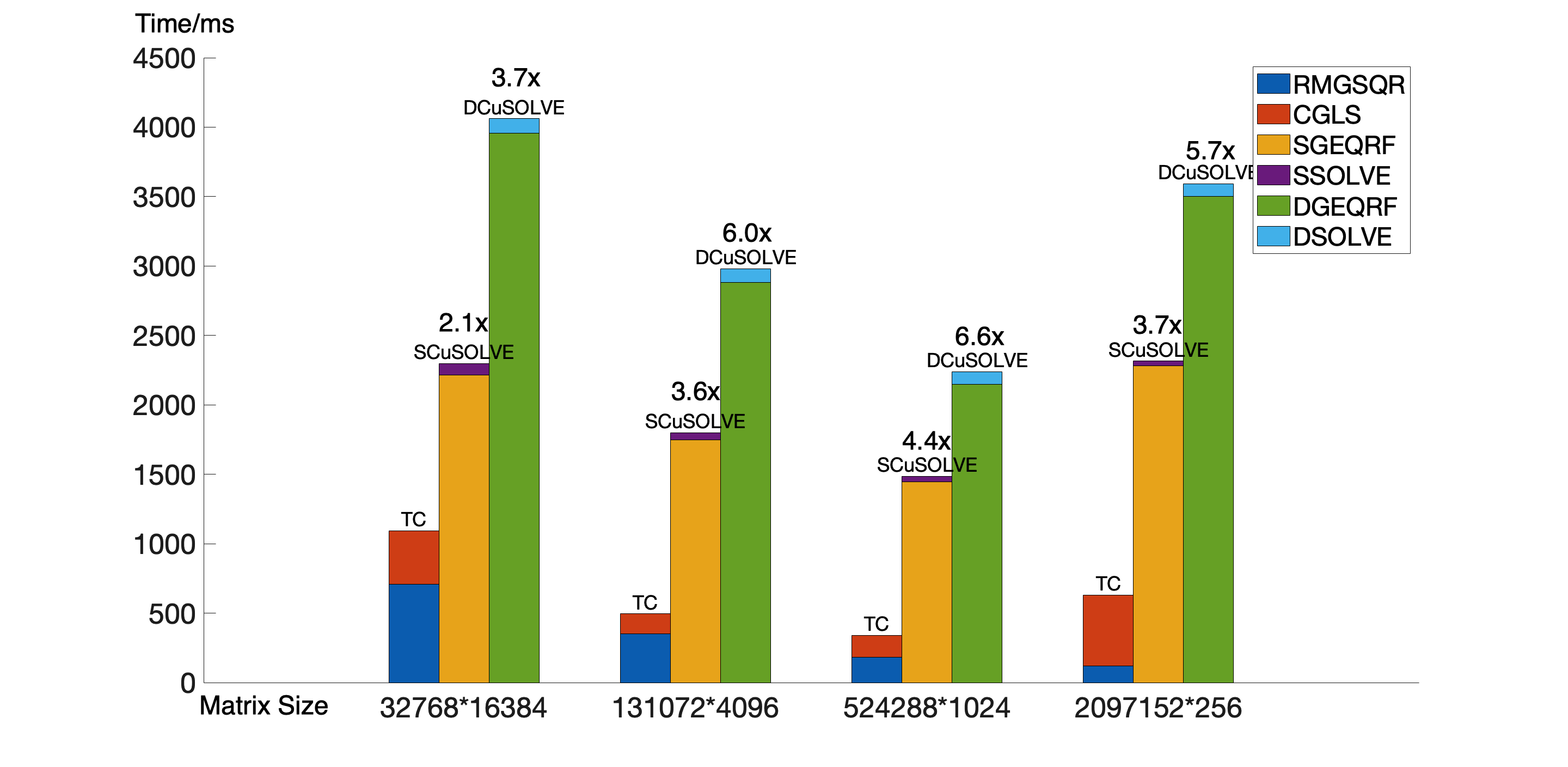}
        \caption{\normalsize Matrix of Type 6: SVD arithmetic distribution with $cond=10^6$}
    \end{subfigure}
    
    \begin{subfigure}[b]{1.0\columnwidth}
        \includegraphics[width=\textwidth]{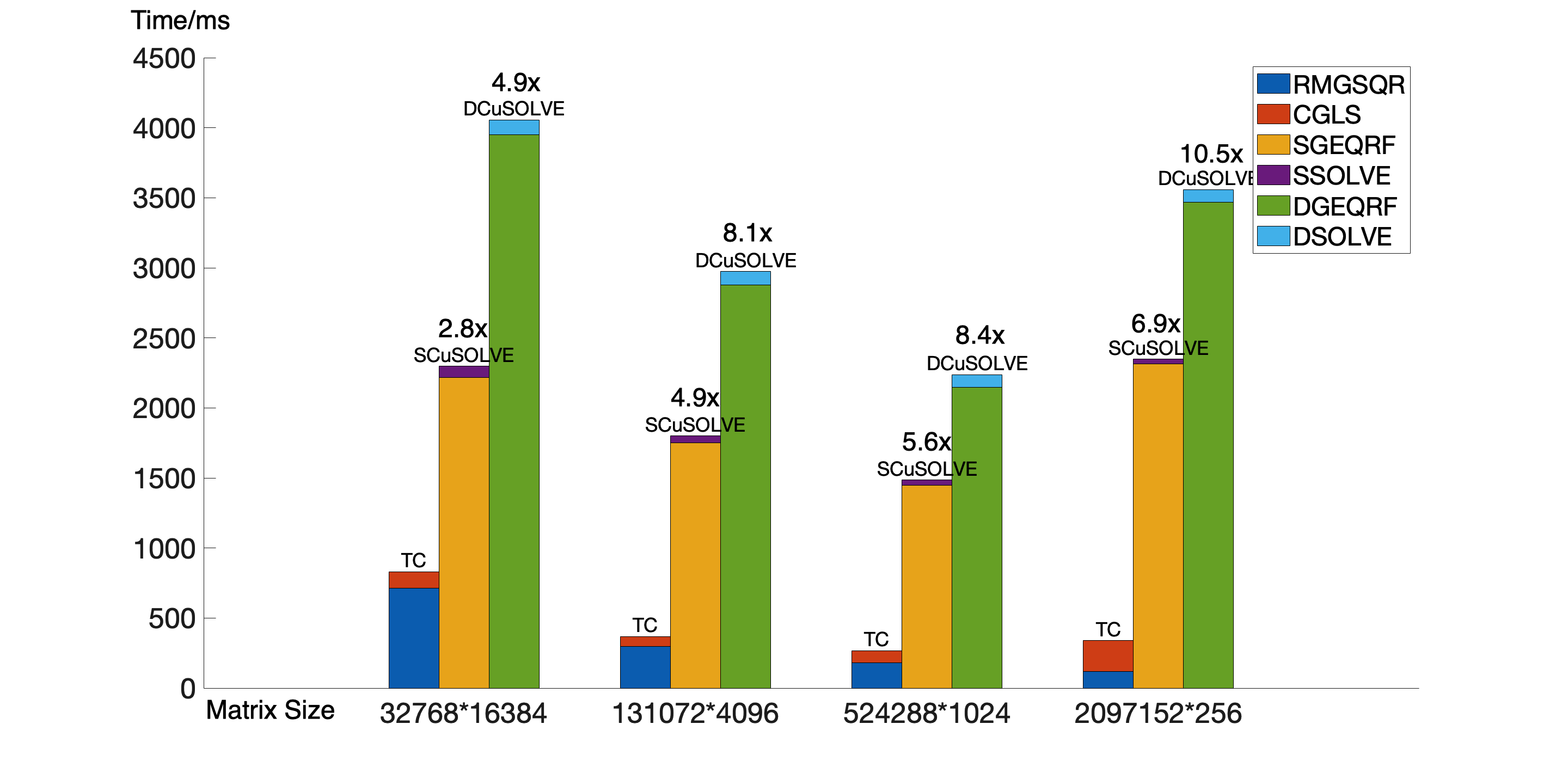}
        \caption{\normalsize Matrix of Type 7: SVD cluster distribution with $cond=10^5$ and  $\sigma_i = \{1,...,1,\frac{1}{cond}\}$}
        \label{fig:3(g)}
    \end{subfigure}
    \begin{subfigure}[b]{1.0\columnwidth}
        \includegraphics[width=\textwidth]{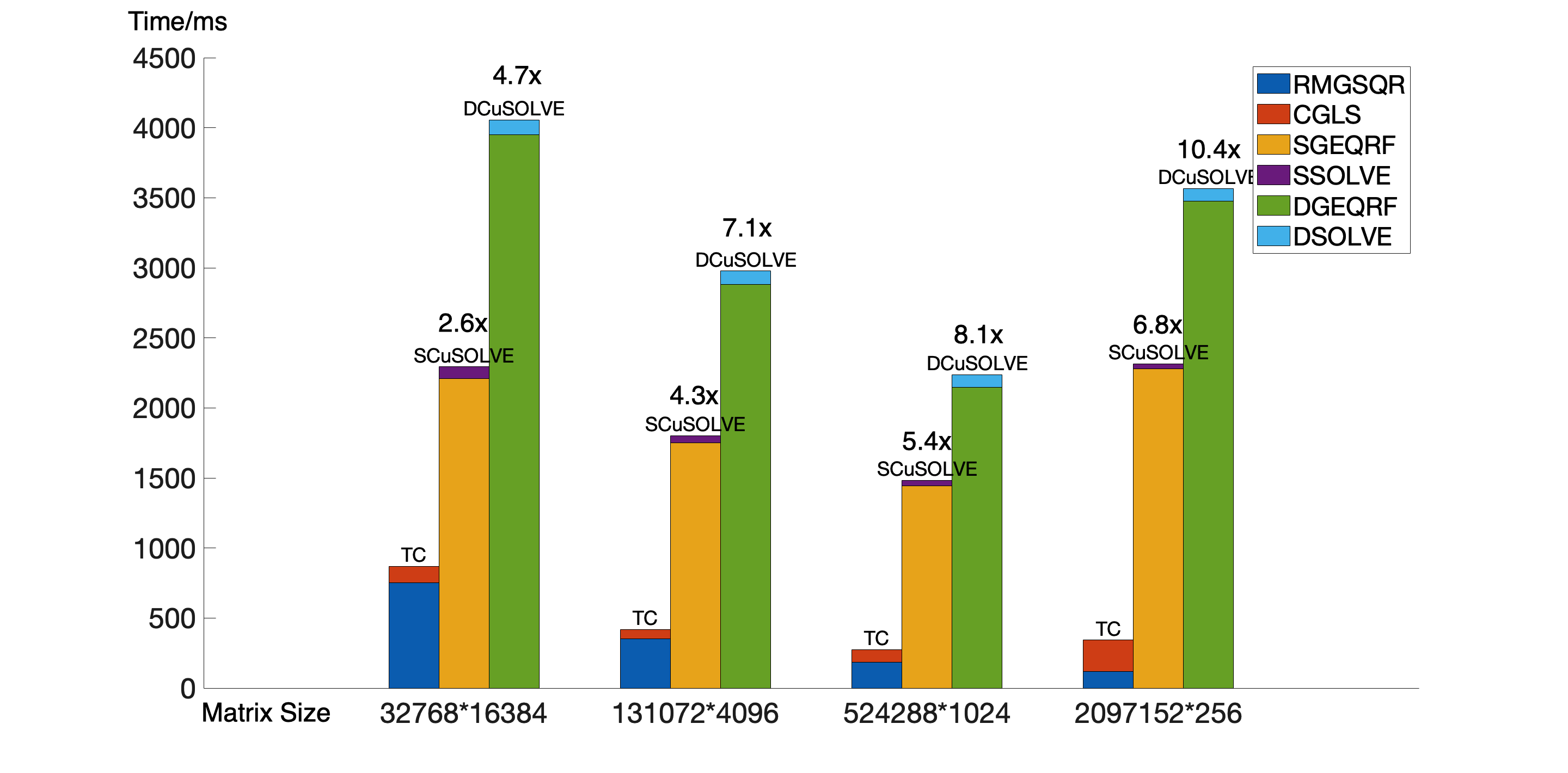}
        \caption{\normalsize Matrix of Type 8: SVD cluster distribution with $cond=10^6$ and  $\sigma_i = \{1,...,1,\frac{1}{cond}\}$}
        \label{fig:3(h)}
    \end{subfigure}
    \newpage
    \caption{Performance in multiples and milliseconds of three linear square problem solvers: RMGSQR iterative solver(left bar), cuSolver SGEQRF direct solver(middle bar) and cuSolver DGEQRF direct solver(right bar) for different matrix types and sizes.}\label{fig:LLS_perf}
\end{figure*}

Based on the performance on QR factorization, we are also expecting a considerable speed up on solving LLS problems. In order to get the same accuracy level with direct LLS solver provided by cuSOLVER, we combine RMGSQR and CGLS together (Algorithm~\ref{alg:cgls_qr}) to 
refine the solution accuracy. 
Figure~\ref{fig:LLS_perf} shows the comparison between time cost of RMGSQR plus CGLS iterative solver and direct solver (SGEQRF+SORMQR+STRSM, see Algorithm~\ref{alg:lls_qr}), note that the RMGSQR solution has \textbf{double precision accuracy}. Obviously, we spend more time in CGLS when compared with direct solvers, which results in somehow a lower speedup than QR factorization. But it is still a tremendous improvement on solving LLS problems. Similarly, there is the some tendency that taller and thinner matrices tend to perform better, which is in line with our imagination from the experiments on QR factorization. 

Generally speaking, CGLS converges pretty fast with preconditioned $AR^{-1}$.  In the case of uniformly random matrix 32768$\times$16384, it can reach a pretty good accuracy in 20 iterations which only costs 300ms. It's relatively slow only if compared with direct solvers. If we regard RMGSQR and CGLS as entirety, it's extremely fast.

However, uniform matrix is typically well-conditioned and it should have a fast converge speed. 
The convergence rate of an iterative solver like CGLS depends strongly on the spectrum property of the matrix $A$. 
To make the LLS problems more general, we generate different types of matrix with different singular value distribution and condition number. We expect results to be condition-distribution-related, that is, the larger condition number the matrix has, the larger number of iterations it will take. In some extreme cases, CGLS cannot converge to satisfied accuracy and we will discuss it in more details next section. Figure~\ref{fig:3(a)} to Figure~\ref{fig:3(h)} illustrates the relationships in terms of condition number, distribution and number of iterations, and it is consistent to our anticipation.

\subsubsection{Accuracy}
\begin{figure}
    \centering
    \includegraphics[width=\columnwidth]{{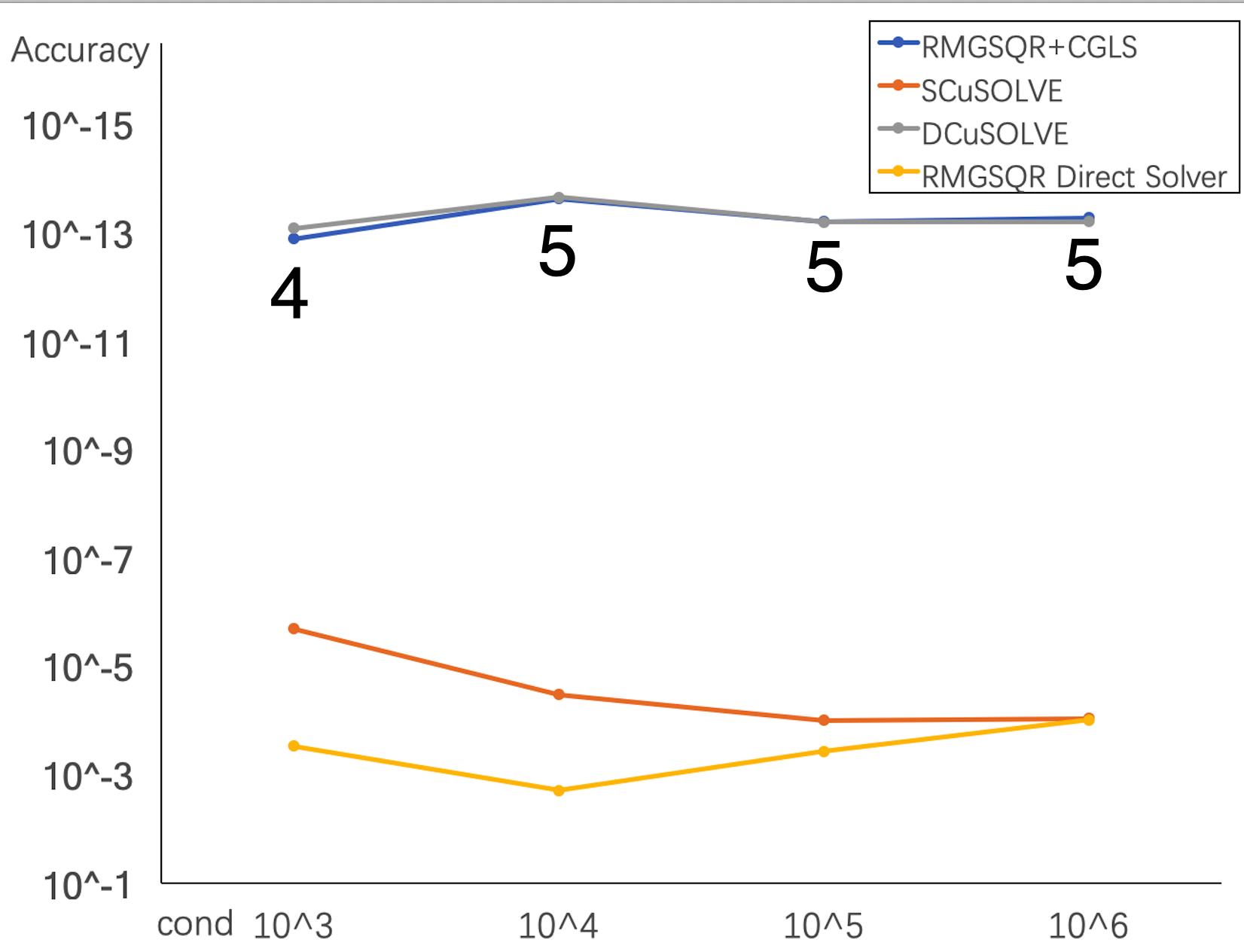}}
    \caption{LLS accuracy: matrix size 32768*16384 with SVD cluster2 distribution, condition number varies from $10^3$ to $10^6$. Comparison between SCuSOLVE, DCuSOLVE,RMGSQR Direct Solver and RMGSQR+CGLS}
    \label{fig:lls_accu}
\end{figure}{}
At first we would like to show the observations on the accuracy based on $x = R^{-1}(Q^Tb)$. Because RMGSQR involves with half precision, so we are not expecting to see as accurate result as cuSOLVER can provide. As the accuracy showed in Fig~\ref{fig:lls_accu}, we can conclude that in most cases, RMGSQR direct solver perform worse than SGEQRF solver and the difference is around two orders of magnitude. It explains why we need iterative methods as safeguard. 

Fig~\ref{fig:lls_accu} also compares DGEQRF direct solver, SGEQRF direct solver and RMGSQR iterative solver accuracy with several condition numbers. For RGEQRF iterative solver, we choose a somehow best tolerance that will give us a relatively accurate result and
reasonable converge speed. We can observe that if the matrix condition is not very bad, RMGSQR and CGLS is able to generate at least the some level of accuracy with DGEQRF direct solver with  small number of iterations(shown by the digits in Fig.\ref{fig:lls_accu}).

To sum up, in terms of accuracy, we can claim that Recursive MGS QR and CGLS iterative method is able to provide a reliable result when compared with double precision Householder QR LLS direct solver. 

\subsubsection{Limitations}
\begin{figure}
    \centering
    \includegraphics[width=\columnwidth]{{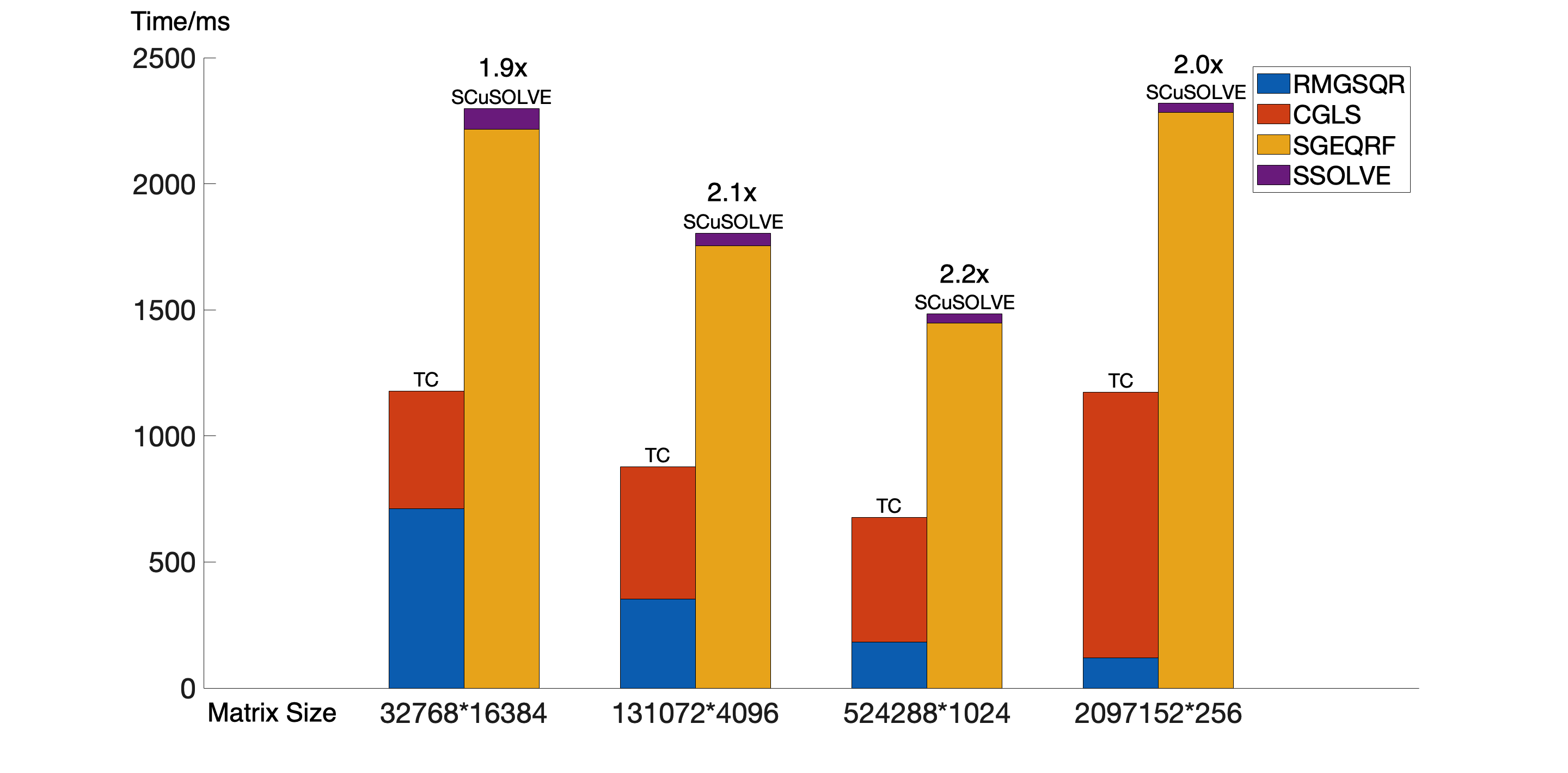}}
    \caption{LLS performace: RMGSQR + CGLS (left bar) vs. cuSolver SGEQRF direct solver (right bar) for different matrix sizes. Matrix type: SVD geometric distribution, $cond = 10^4$ }
    \label{fig:lls_geo}
\end{figure}{}

According to the experiments on SVD geometric distribution(Fig.\ref{fig:3(d)}), we can find the performance on this type of matrix is not as impressive as other types.The reason is that, actually, in this case, CGLS takes 20-30 iterations to converge to $1e-12$(the same accuracy with DCuSOLVE), while other matrix types typically take less than 10 iterations to converge. We also test SVD geometric distribution with $cond=10^4$ and it reveals that for matrix size 32768*16384, it needs 200 iterations, which is the max number of iteration we can tolerate, to converge to $10^{-6}$ and it's probably because of the distribution of singular values.

As a result, we want to see if CGLS can converge to the same accuracy with SCuSOLVE.  Fig~\ref{fig:lls_geo} shows the time cost on SCuSOLVE and RMGSQR+CGLS. We can also observe a good speed-up with RMGSQR. Hence, if single precision is needed, we can enjoy the acceleration by RMGSQR, otherwise we could turn to DCuSOLVE.

To summarize, we are able to claim that we are better than SCuSOLVE in all cases, although there are some hard problems for which we
cannot achieve double precision accuracy efficiently.  
If double precision
accuracy is desired, these problems are best solved using DCuSOLVE. 




\section{Related Work}

NVIDIA introduced TensorCore technology with their Volta architecture~\cite{Nvidia2017} in 2017. 
Resources about NVIDIA TensorCore include detailed micro-architecture analysis and benchmarking
\cite{Jia2018}, an early report on the programmability, performance, and precision~\cite{Markidis2018}.
In~\citep{dakkak_accelerating_2019} important parallel primitives reduction and scan is accelerated with TensorCore.
In~\cite{Haidar2017,haidar_design_2018,Haidar2018a} TensorCore was used for accelerating
linear system solvers in the framework of hybrid CPU/GPU linear algebra package MAGMA~\cite{Dongarra2014a}.
There are numerous use cases of half precision or even lower precision in the
application of neural networks. 

The QR factorization, along with LU and Cholesky factorization form
the one half of important matrix factorizations in numerical linear algebra.
QR factorization can be used to solve linear system, linear least square problems, orthogonalization
of a set of vectors, and eigendecompositions; see the encyclopedic book~\cite{Golub2012}
for more details and pointers.  These factorizations for the core of popular
linear algebra packacges such as LAPACK~\cite{Anderson1999}
and Eigen~\cite{eigenweb} for general CPUs,
PLASMA~\cite{Dongarra2017} on multi-core systems, 
ScaLAPACK~\cite{Blackford1997a} and Elemental~\cite{Poulson2013} for distributed
memory systems, and cuSOLVER\footnote{\url{https://developer.nvidia.com/cusolver}}/cuBLAS\footnote{\url{https://developer.nvidia.com/cublas}} for NVIDIA GPU accelerators as part of CUDA libraries, and SLATE~\cite{kurzak_designing_2017} on distributed heterogeneous CPU/GPU systems.
There are primarily three main algorithms for QR factorization: classic Gram-Schmidt,
modified Gram-Schmidt, and Householder QR~\cite{householder_unitary_1958}. 
See a blog post from Cleve Moler\footnote{\url{https://blogs.mathworks.com/cleve/2016/10/03/householder-reflections-and-the-qr-decomposition/}} for a simple
comparison, and the book~\cite{stewart1998matrix} for details. The
high performance implementation of Householder QR depends
on blocking, i.e. aggregating several Householder reflections
into a single matrix-matrix multiplication. The scheme was
developed in~\cite{schreiber_storage-efficient_1989} and used in virtually all
high performance numerical linear algebra packages. Communication-Avoiding
QR is discussed in~\cite{anderson_communication-avoiding_2011,demmel_communication-optimal_2012}. 

The use of QR factorization as a stable method to solve linear least square problem is standard direct method. Iterative methods
for least square problems are also possible, and may be 
preferred for very large scale and sparse problems. CGLS
appeared in~\cite{hestenes1952methods} together with 
the discovery of Conjugate Gradient method; there's another
mathematically equivalent but numerically more stable one
called LSNR~\cite{paige_lsqr:_1982}. In this paper, we take
a somewhat unsual approach in using iterative method for
a general dense problem. 

The roundoff error analysis of half precision floating point arithmetic
is only emerging. The report \cite{Higham2018} provides some statistical
roundoff error analysis that is more suitable for half precision, as
traditional deterministic analysis is too pessimistic to give any useful
error bound. These papers~\cite{Carson2017a,Carson2017} proposes and
analyzes a mixed half,single, and double precision linear solver.

The closest related work is probably the linear solver based on TensorCore ~\cite{Haidar2017,haidar_design_2018,Haidar2018a}. 
This work shares
some ideas with those recent works in that both compensate the loss of precision from 
TensorCore by combining an iterative solver or iterative refinement.  Both contribute to
the broad effort in bringing TensorCore to linear algebra.  The distinction
is that this paper considers QR factorization instead of LU factorization,
and proposes an GPU only instead of hybrid CPU/GPU.

\section{Conclusion Future Work}
Modern processors and accelerators are beginning to support half precision
(16 bit) arithmetic and special units that does half precision matrix-matrix
multiplication extremely efficiently.  We explored its use in accelerating
the QR factorization, and in solving linear least square problem.  We demonstrate
a way to substantially speedup QR factorization and LLS solving using NVIDIA
TensorCore half precision matrix multiplication while achieving double precision
accuracy. 

Future work include extension to non-linear least square, least square problems
with constraints, and under-determined least square problem, etc.  

\bibliography{references}
\end{document}